\documentstyle[twoside,fleqn,espcrc2]{article}

\newcommand{\AmS}{{\protect\the\textfont2
  A\kern-.1667em\lower.5ex\hbox{M}\kern-.125emS}}

\title{Lattice Gauge Field Interpolation for Chiral Gauge Theories}

\author{Pilar Hernandez and Raman Sundrum\thanks{Research supported by
NSF-PHYS-92-18167} ~~~~~~~~~~
~~~~~~~~~~~~~~~~~~~~~~~~~~~~~~~~~~~~~~~~~~~~~~~~~~~~~~~~~~~~~~~
~~~~~~~~~~~~~~~~~~~~~~~~~~~~~~~~~~~~~~~~~~~~~~~~~~~~~~~~~~~~~~~~~~~~~~~~
~~~~~~~~~~~~~~~~~~
Lyman Laboratory of Physics, Harvard University, Cambridge, MA 02138, USA}

\begin{document}

\pagestyle{empty}

\begin{abstract}
The importance of lattice gauge field
interpolation for our recent non-perturbative 
formulation of chiral gauge theory is emphasized. We illustrate how the
requisite properties are satisfied by our recent four-dimensional
non-abelian interpolation scheme, by going through the simpler case of
$U(1)$ gauge fields in two dimensions. 
\end{abstract}

\maketitle

\section{INTRODUCTION}

In every major scenario for physics above the $TeV$ scale
non-perturbative chiral gauge theory dynamics is expected to play an important
role, yet our understanding of this dynamics is very limited. Recently we
proposed a non-perturbative formulation of chiral gauge theory  on
the lattice, in the hope that the important features of these theories will be
calculable in future computer simulations \cite{us,int}. 
We briefly review it here,
focussing on a lattice gauge field interpolation procedure which is the 
crucial feature of our construction.

The fermions are taken to live on a euclidean lattice with spacing $f$.
Fermion doublers are eliminated by the Rome Group method of using a gauge
non-invariant Wilson term \cite{rome}. (The Nielsen-Ninomiya
Theorem  tells us that we must break chiral gauge invariance to
eliminate the fermions doublers.)
  The central problem is then to recover
gauge invariance in the continuum limit. Before summing over gauge fields
this can be arranged  in a fairly simple way for any $anomaly$-$free$
theory \cite{gsbk,us}. 
However, 
once gauge fields are integrated over, new
 divergences can lead to uncontrollable violations of gauge invariance. A
solution to this problem is to cut off the   gauge
field momenta by a scale $\Lambda_b \ll 1/f$ ($f\Lambda_b, f, 1/\Lambda_b 
\rightarrow 0$ in
the continuum limit.) 
 Ref. \cite{us} gives detailed lattice power-counting
arguments to show that this can be achieved by obtaining the $f$-lattice
gauge fields as an interpolation of gauge fields living on a lattice of
spacing $b \sim 1/\Lambda_b \gg f$, which are summed over using the standard
Wilson action. In practice it is possible that $b/f$ may not have to be too
large for computing the properties of low-lying states \cite{us}. (For another
possible way of cutting off gauge field momenta see ref.
\cite{slavnov}.)

It is sufficient for the interpolation procedure to satisfy the following
properties \cite{us}, 
most easily stated by imagining interpolating the $b$-lattice
link variables, $U$, all the way to
the continuum to give gauge fields, $a_{\mu}[U]$.  (i) Transverse continuity: 
the interpolation describes a differentiable continuum gauge field $inside$
each $b$-lattice hypercube, whose transverse components are continuous
across hypercube boundaries (the longitudinal components can jump). 
(ii) Lattice spacetime symmetries should
be respected. (iii) Gauge covariance: 
A $b$-lattice gauge transformation changes the
interpolation only by a continuous gauge transformation. 
  (iv) Locality: The
gauge-invariant behavior of the interpolation should depend locally 
on $U$, in particular it is sufficient if the trace of any continuum
Wilson loop depends only on the $U$ lying on $b$-hypercubes through which
the Wilson loop passes.  

We have detailed such an interpolation procedure for non-abelian gauge
fields in four dimensions in ref. \cite{int}. Below, we describe the more 
transparent case of interpolating $U(1)$ gauge fields in $two$ dimensions,
 following a procedure which readily generalizes to the non-abelian
case. For simplicity we deal with 
 the case $f = 0$ (the continuum) and work in units where $b = 1$.

\section{$U(1)$ 2-D Interpolation}

In order to maintain transverse continuity it is helpful to build the
interpolation from the lowest dimensional sublattices up. We therefore
begin by interpolating the link variables,
\begin{equation}
U_{\mu}(s) = e^{i A_{\mu}(s)}, ~~|A_{\mu}(s)| < \pi, 
\end{equation}
along the points of each plaquette edge.
(We are neglecting the measure-zero set of lattice fields where at
least one of the link variables equals exactly $-1$.) The simplest such
interpolation is
\begin{equation}
a_{\mu}(s + t \hat{\mu}) \equiv A_{\mu}(s),  ~~0 \leq t < 1.
\end{equation}
Note that parallel transport along the links agrees between the lattice and
the continuum fields.

We now $attempt$ to interpolate the lattice field into a plaquette interior
in
such a way as to agree with the above interpolation on the plaquette edges.
In order to satisfy locality we try to do this interpolation for each
plaquette, using as input only its bounding link variables. 
In order to satisfy gauge covariance the strategy is to do a $lattice$
gauge transformation on the bounding links of the plaquette which put the
link variables into a complete axial gauge. Thus all gauge equivalent
lattice fields on the plaquette edges are taken to the same gauge-fixed 
field (`almost' the same in the non-abelian case), 
which we denote by $\overline{U}$. 
This lattice configuration will then be smoothly interpolated to the plaquette
interior to give a continuum field $\overline{a}_{\mu}$. We will then try to 
find a smooth gauge transformation inside the plaquette which makes
the result agree with the one-dimensional edge interpolation, eq. (2).
At this last stage we will fail, but in a way which we can understand and
then correct.

In detail, let us fix some plaquette and use $local$ coordinates $(z_1,
z_2), z_{\mu} = 0, 1$ for the vertices of the plaquette. The lattice
gauge transformation,
\begin{equation}
\Omega[U](z) = U_1(s)^{z_1} U_2(s + z_1 \hat{1})^{z_2},   
\end{equation}
takes $U_{\mu}$ to 
\begin{eqnarray}
\overline{U}_1(0,1) &=&  U_2(s) U_1(s + \hat{2}) U_2^{-1}(s + \hat{1})
U_1^{-1}(s), \nonumber \\
\overline{U}_1(0,0) &=& \overline{U}_2(0,0) = \overline{U}_2(1,0) = 1. 
\end{eqnarray}
 This lattice field is easily interpolated into the plaquette interior,
\begin{eqnarray}
\overline{a}_1(t_1, t_2) &=&  t_2 \overline{A}_1(0,1) \nonumber \\
\overline{a}_2(t_1, t_2) &=& 0,  
\end{eqnarray}
where $(t_1, t_2): 0 \leq t_{1,2} \leq 1$
 are local continuum coordinates for the plaquette interior.

The problem is now to find a $continuum$ gauge transformation $\omega$ 
which takes 
$\overline{a}_{\mu}$ to a gauge field agreeing  with  eq. (2),
so that we can be assured of transverse continuity across plaquette
boundaries. In fact it is not hard to see that this demand essentially
fixes $\omega$ on the plaquette boundary to be
\begin{eqnarray}
\omega(t_1, 0) &=& e^{i t_1 A_{1}(s)}, \nonumber \\
\omega(1, t_2) &=& U_1(s) e^{i t_2 
A_{2}(s + \hat{1})}, 
\nonumber \\
\omega(0, t_2) &=& e^{i t_2 A_{2}(s)}, \nonumber \\
\omega(t_1, 1) &=& 
U_2(s) e^{i t_1 (A_{1}(s + \hat{2}) - \overline{A}_{1}(0,1))},
\end{eqnarray}
thereby specifying a map from the plaquette boundary (topologically a
circle) to $U(1)$ (topologically also a circle). We can therefore associate
a topological winding number (integer) to this map for each plaquette.
Unless this winding is zero, $\omega$ cannot be continuously extended from
the plaquette boundary to the interior, and we are stuck. It is simple to
show that the winding number associated
to the plaquette at $s$, $N(s)$,  is equal to 
\begin{equation}  
I[\frac{A_{2}(s) + A_{1}(s+\hat{2}) - A_{2}(s+\hat{1}) -
A_{1}(s)}{2 \pi}], 
\end{equation}
($I[y] \equiv$ nearest integer to $y$) 
and is generically non-zero. 

The way out of  this impasse is to generalize the
edge interpolation to 
\begin{equation}
a_{\mu}(s + t \hat{\mu}) = A_{\mu}(s) + 2 \pi n_{\mu}(s),  ~~0 \leq t < 1,
\end{equation}
where the $n_{\mu}(s)$ are integer-valued and do not affect agreement of
parallel transport between the link variables and the interpolation. (In the
notation of ref. \cite{int}, $n_{\mu} = -\epsilon_{\mu \nu} N_{\nu}$.) 
If these
integers are chosen to satisfy
\begin{equation}
\sum_{\mu \nu} \epsilon_{\mu \nu} (n_{\nu}(s + \hat{\mu}) - n_{\nu}(s)) = N(s),
\end{equation}
it is easy to
show that this new definition differs from the original by a gauge
transformation defined on the plaquette boundary with winding number
$-N(s)$. Therefore  the gauge transformation which makes
$\overline{a}_{\mu}$ agree with eq. (9),
$\tilde{\omega}$, has winding number zero and can be smoothly extended to
the plaquette interior, allowing us to get an interpolation $a_{\mu} =
\overline{a}^{\tilde{\omega}}_{\mu}$. A simple choice for this extension of
$\tilde{\omega}$ yields for $a_{\mu}(s_1 + t_1, s_2 + t_2)$, 
\begin{eqnarray}
a_1 &=& (1 -\; t_2)\; ( A_1(s) + 2 \pi n_1(s) ) \nonumber\\ 
&+& t_2\; ( A_1(s+\hat{2}) + 2 \pi n_1(s+\hat{2}))\nonumber\\
a_2 &=& (1 -\; t_1) \;( A_2(s) + 2 \pi n_2(s) ) \nonumber\\
&+& t_1 \; ( A_2(s+\hat{1}) + 2 \pi n_2(s+\hat{1}) ). 
\label{interqed}
\end{eqnarray}

If $\sum_s N(s) \neq 0$ then in fact there is no consistent
solution to eq. (9). The reason is that the boundary conditions for the
interpolation then correspond to a continuum configuration
 with topological charge $\sum_s N(s) \neq 0$, which cannot be represented
by a single smooth periodic gauge field. While configurations with non-zero
topological charge are physically important, we do not need them in our
proposal for lattice chiral gauge theory, because their effects can be
inferred from the sector with zero topological charge using cluster
decompostion for the full theory. For $\sum_s N(s) = 0$ eq. (9) has many
solutions and it is simple to pick one \cite{int}.

Let us check the four central requirements for a successful interpolation.
(i) It is straightforward to see that eq. (10) defines a transversely
continuous gauge field. (ii) Even though we had to pick the axes of our
complete gauge fixing somehow, the $gauge$-$invariant$ behavior of our interpolation is
covariant under lattice translations and rotations, though the gauge
dependent form is not. To see this in the $U(1)$ case is easy, since from
eqs. (10, 9) the continuum field strength is a constant in each plaquette
and is just the logarithm of the plaquette
field strength (with absolute value less than $\pi$). 
For the same reason (iii) and (iv) are also obvious, the non-locality in choosing the
$n_{\mu}$ does not infect the gauge invariant part of the interpolation
(ie. the field strength for the $U(1)$ case). 

\subsection{Non-abelian 4-D Interpolation}

While non-abelian interpolation in four dimensions is technically more
complicated, the basic steps
are the same. The topological obstruction to making higher dimensional
interpolations agree with lower dimensional ones now occurs in
four dimensions, because the smallest non-abelian group, $SU(2)$, is
topologically the 3-sphere as is a hypercube boundary. The resolution of
the problem generalizes the 2-D $U(1)$ case. One extra complication not
seen in two dimensions is that the choice of axes for the complete gauge
fixing (important for maintaining gauge covariance)
 $does$  lead to a breaking of lattice rotational covariance in the 
 gauge-invariant behavior of the interpolation. To repair this one needs to
allow the orientation of the axes in each hypercube to be different and to
determine this orientation by the gauge-invariant behavior of the link
field $U$ itself. Then if $U$ is rotated, so do the axes. See ref.
\cite{int}.

 The interpolation
in ref. \cite{kron}  differs in that  
the authors directly interpolate $\omega$ 
instead of $\tilde{\omega}$, thus getting singular gauge fields whenever 
any $N(s)$ is non-zero. Such singular gauge fields are unsuitable for our
formulation of chiral gauge theories. Their interpolation also breaks lattice
rotational covariance in the non-abelian case.

\end{document}